\documentclass[10pt, a4paper]{article}

\usepackage[]{lrec-coling2024} 
\usepackage{times}
\usepackage{latexsym}
\usepackage{graphicx}
\usepackage{xcolor}
\usepackage{array}
\usepackage{caption}
\usepackage{subcaption}
\usepackage{amsmath}
\usepackage{amsfonts}
\usepackage{amssymb}
\usepackage{bbm}
\usepackage{multirow}
\usepackage{subcaption}
\newcommand\scalemath[2]{\scalebox{#1}{\mbox{\ensuremath{\displaystyle #2}}}}

\title{CuSINeS: Curriculum-driven Structure Induced Negative Sampling for Statutory Article Retrieval}

\name{Santosh T.Y.S.S, Kristina Kaiser, Matthias Grabmair} 

\address{School of Computation, Information, and Technology; \\ Technical University of Munich, Germany
\\
\{santosh.tokala, kristina.kaiser, matthias.grabmair\}@tum.de\\}

\abstract{
In this paper, we introduce CuSINeS, a negative sampling approach to enhance the performance of Statutory Article Retrieval (SAR). CuSINeS offers three key contributions. Firstly, it employs a curriculum-based negative sampling strategy guiding the model to focus on easier negatives initially and progressively tackle more difficult ones. Secondly, it leverages the hierarchical and sequential information derived from the structural organization of statutes to evaluate the difficulty of samples. Lastly, it introduces a dynamic semantic difficulty assessment using the being-trained model itself, surpassing conventional static methods like BM25, adapting the negatives to the model's evolving competence. Experimental results on a real-world expert-annotated SAR dataset validate the effectiveness of CuSINeS across four different baselines, demonstrating its versatility.
 \\ \newline  }

\begin{document}

\maketitleabstract

\section{Introduction}
In an age marked by complex legal challenges, there's a growing imperative to bridge the gap between legal expertise and public comprehension 
\cite{ponce2019global}. Statutory article retrieval (SAR) involves finding relevant statutes for a legal question and is a vital initial step in legal assistance. 
Traditionally, SAR methods have been explored using the COLIEE Statute Law Corpus \cite{rabelo2021coliee}, containing questions linked to relevant articles from the Japanese Civil Code. However, these questions which are obtained from legal bar exam yes or no questions, are quite different from those posed by ordinary citizens, often being vague and underspecified. To address this, \citealt{louis2022statutory} developed the Belgian Statutory Article Retrieval Dataset (BSARD), featuring french legal questions from Belgian citizens labeled by legal experts with references to relevant articles from Belgian legislation, 
which we use in our study.

Traditional SAR techniques included BM25, TF-IDF \cite{yoshioka2018overview}, Indri \cite{strohman2005indri} and Word Movers’ Distance \cite{kusner2015word}. With the rise of pre-trained models, BERT and their ensembles have become popular \cite{kim2019statute, rabelo2021coliee, rabelo2022overview}. Recently, dense retrieval methods have gained attention \cite{louis2022statutory} and were enhanced further through synthetic query generation and legal domain-oriented pre-training \cite{louis2023finding}. Graph neural networks have been applied to enrich article representations by exploiting the interdependencies among articles within the topological structure of legislation, which consist of a network of interconnected statutes organized into codes, books, titles, chapters, and sections, forming a hierarchical and sequential framework.

Despite these improvements, 
a key aspect has been overlooked: 
how to construct high-quality negative samples for training SAR models. 
Prior methods have primarily relied on BM25-based semantic similarity 
to derive hard negatives. However, there have been no explicit efforts to utilize the structural organization of legislation for mining hard negatives. While \citealt{louis2023finding} used this structural information to derive article representations, our approach utilizes this to mine hard negatives, which is orthogonal to their work.

This structural organization reveals that the distant statutes, with greater shortest path, 
cover broader legal themes while the statutes with lesser shortest path deal with similar legal concepts. This makes  negative articles distant from the candidate positive easier to distinguish, while the near ones  are
more difficult to distinguish. 
This insight into difficulty estimation based on structure complements the traditional semantic approach. 
Additionally, we enhance semantic difficulty estimation by dynamically assessing it with the retrieval model being trained. This goes beyond the static estimation derived from BM25, which is model-independent. This makes the model expose to those negatives during training based on its current competence.

Furthermore, inspired by 
curriculum learning, which suggests that learning often begins with simpler samples before gradually moving to more complex ones \cite{bengio2009curriculum}, we introduce a curriculum-based scheduling of negative samples. This approach guides the model to focus on differentiating positive articles from easier negatives in the initial learning stages and gradually transition to learning sophisticated reasoning from difficult negatives as the training advances. This helps the model find better local minima 
by mitigating the negative impact of difficult samples in the early stages of training \cite{hacohen2019power}. 

Combining these three insights, we introduce CuSINeS, a Curriculum-driven Structure Induced Negative Sampling approach, which is model-agnostic and can be employed with training any SAR model. We apply CuSINeS on top of four SAR models on the BSARD dataset showcasing the versatility of our approach. 

\section{Preliminaries}

\noindent \textbf{Statutory Article Retrieval:} 
Given a question $q$ and corpus of statues $S =\{s_1, s_2, \ldots, s_m\}$, the task of SAR is to retrieve a smaller set of statutes $S_q$ ($|S_q|<< |S|)$ ranked in terms of their relevancy to answer the query. We mainly deal with optimizing the recall of the SAR system acting as pre-fetcher, leaving the re-ranker component 
optimized for precision, for future. 
\newline

\noindent \textbf{Dense Retrieval (DR):} They use a dual-encoder architecture \cite{karpukhin2020dense}, where the relevance score is computed using dot product between encodings of query $q$ and statute $s_i$ as 
    $f(q,s_i)= E_q(q) \cdot E_s({s_i})$
where $E_q$, $E_s$ denote query and statute encoder to map each of them into a k-dimensional dense vector  respectively.
\newline

\noindent \textbf{Training with Negative Sampling:} DR models are trained with contrastive loss whose objective is to pull the representations of the query $q$ and relevant articles $S_q$ together (as positives), while pushing apart irrelevant ones $S_q' = S - S_q$ (as negatives) \cite{lee2019latent}. However, 
training with all the negatives is inscalable given larger corpus. To alleviate this issue, negative sampling has been employed where some irrelevant documents are sampled for each query during training making the final objective function as follows:
\begin{equation}
\scalemath{0.8}{
    L(q, S_q, S_q') = \sum_{p \in S_q}  -log  \frac{\exp(f(q,p)/\tau)}{ \sum_{c \in \{p\} \cup S_q'} \exp(f(q,c)/\tau)}}
\end{equation}
where hyperparameter $\tau$ is a scalar temperature.

\section{Our Method: CuSINeS}
Unlike previous approaches that rely solely on hard negatives obtained through BM25 based on semantic relevance and employ them from the initial training stage \cite{louis2022statutory,louis2023finding}, CuSINeS introduce a curriculum-based scheduler that exposes the model to easier negative samples before gradually introducing more challenging ones, facilitating the model to learn over time emulating the human learning process. Further, CuSINeS incorporates structural information to derive difficulty ranking in addition to the semantic one. Additionally, semantic-based ranking is updated dynamically using the model under training, based on its current competence.

\subsection{Difficulty ranking of negatives}
\noindent \textbf{Semantic-based ranking}  
We dynamically compute semantic difficulty of negative articles by assessing their semantic relevance to the query. A higher relevance score indicates a more difficult negative article. This dynamic ranking provides a more nuanced comprehension of the model's learning dynamics. While the model undergoes updates with each mini-batch iteration, we opt to refresh the difficulty rankings at each epoch to reduce the inference cost associated with continuous updates.
\newline

\noindent \textbf{Structure-based ranking} We leverage the statute structure 
to derive the difficulty ranks for negative articles. We consider two views based on structure: (i) Hierarchical view: We determine the difficulty of each negative article by measuring its proximity to the set of positive articles within the hierarchical graph. To create a ranked list of negatives for a given query, we follow these steps: First, we compute the shortest path distance between each positive article and every negative article within the hierarchical graph. Next, we determine the final distance for each negative article by selecting the minimum distance among all the distances to the positive articles. We rank these negatives based on distances, where a lower distance indicates a more difficult negative. (ii) Sequential View: Similar to the hierarchical view, the sequential view treats statutes as a linearized sequence. It calculates the distance score by considering the relative positional information between positive and negative articles in the sequential enumeration of articles and the difficulty rank is obtained similar to hierarchical view. 
\newline 

\noindent \textbf{Combining multiple difficulty ranks} While the semantic difficulty captures the interplay between queries and negative articles, the structural one reflects the relationship between positive and negative articles
, highlighting their complementary nature. We unify these three sets of rankings through reciprocal rank fusion (RRF) \cite{cormack2009reciprocal}. For a query $q$ and its corresponding set of negative articles $S_q'$, we generate rankings $R$ using three methods (each providing a permutation on $1,\ldots,|S_q'|$). We then calculate the RRF score for each negative article and sort them to obtain the cumulative difficulty rankings.

\begin{table}[ht]
\scalebox{0.8}{
\begin{tabular}{|l|l|ccc|c|c|}
\hline
\multirow{2}{*}{\textbf{Method}}                                      & \multirow{2}{*}{\textbf{}} & \multicolumn{3}{c|}{\textbf{R@}}                                                     & \multicolumn{1}{l|}{\multirow{2}{*}{\textbf{MAP}}} & \multicolumn{1}{l|}{\multirow{2}{*}{\textbf{MRP}}} \\ \cline{3-5}
 &                            & \multicolumn{1}{c|}{\textbf{100}} & \multicolumn{1}{c|}{\textbf{200}} & \textbf{500} & \multicolumn{1}{l|}{}                              & \multicolumn{1}{l|}{}                              \\ \hline
BM25                                                                  & Baseline                   & \multicolumn{1}{c|}{49.3}         & \multicolumn{1}{c|}{57.3}         & 63           & 16.8                                               & 13.6                                               \\ \hline
\multirow{2}{*}{DR CB}                                                & Baseline                   & \multicolumn{1}{c|}{77.1}         & \multicolumn{1}{c|}{81.8}         & 86.7         & 35.6                                               & 28.8                 \\
& CuSINeS                     & \multicolumn{1}{c|}{82.6}         & \multicolumn{1}{c|}{86.6}         & 91.6         & 38                                                 & 29.1                                               \\ \hline
\multirow{2}{*}{\begin{tabular}[c]{@{}l@{}}DR+GNN\\ CB\end{tabular}}  & Baseline                   & \multicolumn{1}{c|}{80.2}         & \multicolumn{1}{c|}{83.2}         & 88.6         & 39.2                                               & 32.6                                               \\ 
& CuSINeS                     & \multicolumn{1}{c|}{83.2}         & \multicolumn{1}{c|}{88.1}         & 92.6         & 42.2                                               & 33.4                                               \\ \hline
\multirow{2}{*}{DR LCB}                                               & Baseline                   & \multicolumn{1}{c|}{79.8}         & \multicolumn{1}{c|}{83.9}         & 88.9         & 39.5                                               & 31.3                                               \\ 
& CuSINeS                     & \multicolumn{1}{c|}{83.7}         & \multicolumn{1}{c|}{87.5}         & 92.3         & 41.2                                               & 32.1                                               \\ \hline
\multirow{2}{*}{\begin{tabular}[c]{@{}l@{}}DR+GNN\\ LCB\end{tabular}} & Baseline                   & \multicolumn{1}{c|}{82.6}         & \multicolumn{1}{c|}{85.6}         & 90.1         & 44.6                                               & 35.8                                               \\ 
& CuSINeS                     & \multicolumn{1}{c|}{84.9}         & \multicolumn{1}{c|}{89.6}         & 93.3         & 46.2                                               & 36.2                                               \\ \hline
\end{tabular}}
\caption{Comparison of CuSINeS with \small{Baseline} negative sampling strategy on four dense models. (L)CB denote (Legal)CamemBERT as encoder model. 
}
\label{main-results}
\end{table}

\subsection{Curriculum Scheduler}
Based on cumulative difficulty ranking, we categorize these negatives into various difficulty-level buckets, ranging from easy to difficult. During the training process, we draw negative samples for each query from all the buckets. In initial iterations, a larger proportion of samples come from the easier buckets with smaller proportion from the difficult ones. As training progresses, the ratio gradually shifts, allocating a higher share of difficult samples in subsequent iterations. This adaptive scheduling enhances the model's ability to learn from a range of difficulty examples, akin to the Leitner system of spaced repetition that improves human learning.

\section{Experiments}
\subsection{Dataset \& Baselines}
We use BSARD \cite{louis2022statutory} containing  1108 french legal questions, with references to relevant articles from a corpus of 22,600 Belgian legal articles. A query can have multiple relevant legal articles. 
\newline

\noindent \textbf{Baselines} We derive following baselines from \citealt{louis2022statutory,louis2023finding} (i) BM25 \cite{robertson1995okapi} as sparse retrieval baseline. (ii) Dense Retrieval (DR) model where BERT model is used as query encoder and hierarchical version of BERT is used as article encoder to account for longer length of articles. This hierarchical version splits longer text into various segments, obtains [CLS] representation for each segment and passes them through another transformer layer to obtain the final representation through max pooling. (iii) DR+GNN where dense retrieval model is augmented with graph attention network, a variant of graph neural network, to enrich article representations by fusing information from a legislative graph constructed from hierarchical organization of statutes. We experiment with two initializations in each of the dense models: one with the French CamemBERT \cite{martin2020camembert} and other with the LegalCamemBERT \cite{louis2023finding} which is further pre-trained on BSARD corpus. All these baselines employ BM25 for negative mining with a fixed training schedule i.e. use these hard negatives from the initial training stage. We apply our negative sampling method, CuSINeS, to these four dense models. CuSINeS is model-agnostic and can be integrated into the training of any model. 

\subsection{Implementation Details}
We adhere to the baseline configuration outlined in previous work by \citealt{louis2023finding}. In our hierarchical article encoder, we initialize the second-level encoder with a two-layer transformer encoder featuring a hidden dimension of 768, an intermediate dimension of 3072, 12 heads, 0.1 dropout rate, and the GeLU non-linearity function. Our training process for DR spans 15 epochs with a batch size of 24, employing AdamW optimizer \cite{loshchilov2018decoupled} with hyperparameters $\beta_1$ = 0.9, $\beta_2$ = 0.999, $\epsilon$ = 1e-7, a weight decay of 0.01, and a learning rate warm-up for the first 5\% of training steps, reaching a maximum value of 2e-5, after which linear decay is applied. For GNN, we conduct 25 epochs of training with a batch size of 512, using the AdamW optimizer with a learning rate of 2e-4. The best model is determined based on performance evaluation on the validation set. For our adaptive curriculum strategy, we sample 0.7\%, 0.2\%, 0.1\% from easy to difficult buckets in the initial 5 epochs, 0.15\%, 0.7\%, 0.15\% in the next 5 epochs and 0.1\%, 0.2\%. 0.7\% in the last 5 epochs.

\begin{figure*}
    \centering
    \begin{subfigure}{0.22\textwidth}
        \includegraphics[width=\linewidth]{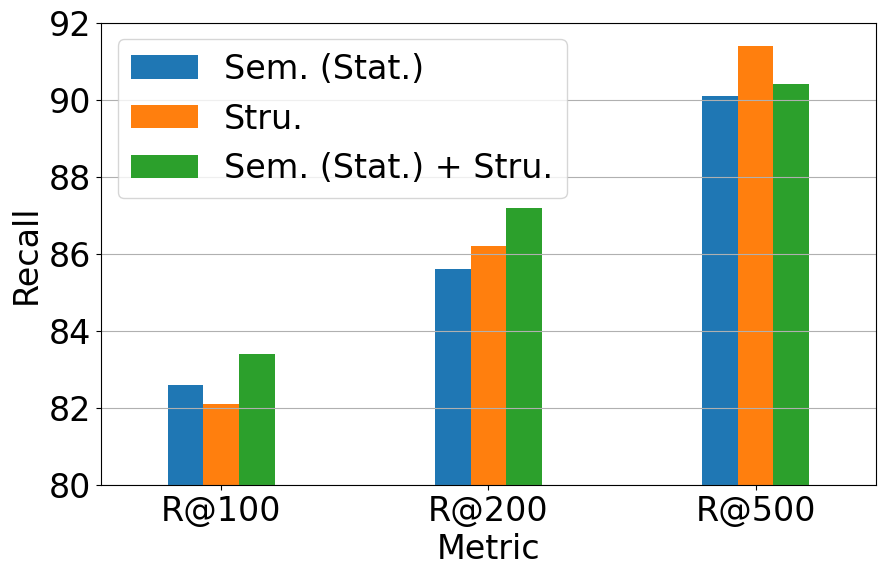}
        \caption{Structural information in negative mining}
        \label{strcture-fig}
    \end{subfigure}
    \hfill 
    \begin{subfigure}{0.22\textwidth}
        \includegraphics[width=\linewidth]{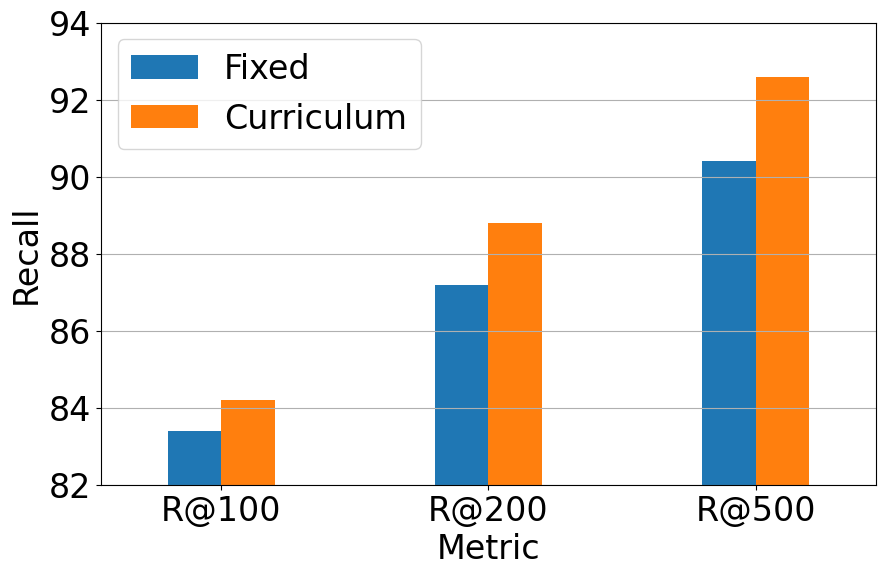}
        \caption{Training scheduler: Fixed vs Curriculum}
        \label{scheduler-fig}
    \end{subfigure}
    \hfill
    \begin{subfigure}{0.22\textwidth}
        \includegraphics[width=\linewidth]{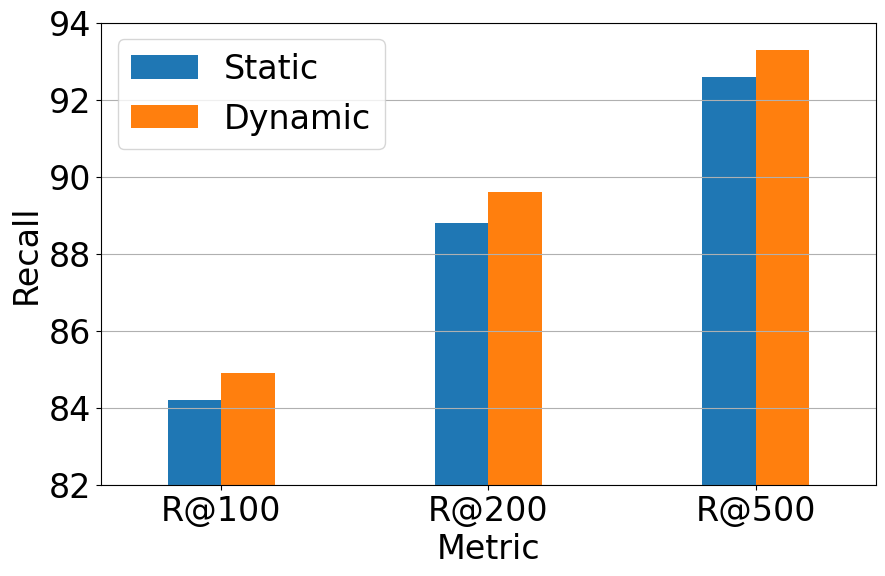}
        \caption{Semantic difficulty ranking: Static vs Dynamic}
        \label{semantic-fig}
    \end{subfigure}
    \hfill
    \begin{subfigure}{0.22\textwidth}
        \includegraphics[width=\linewidth]{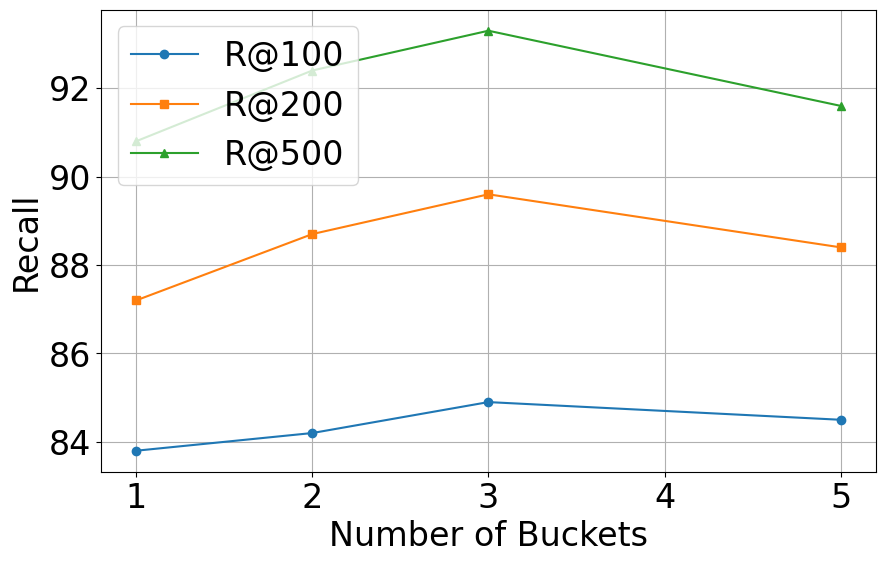}
        \caption{Number of Buckets for curriculum scheduling}
        \label{buckets-fig}
    \end{subfigure}
    \caption{Analysis of sub-components of CuSINeS}
\end{figure*}

\subsection{Performance comparison}
Following previous work, we evaluate the retriever's performance using Recall@k (R@k) (k=100,200,500), Mean Average Precision (MAP) and Mean R-Precision (MRP). R@K measures the proportion of relevant articles in the top-k candidates, with results averaged across all instances. MAP and MRP provide the mean of average precision and R-Precision scores for each query where average precision is the average of Precision@k scores for every rank position of each relevant document and Precision@k represents the proportion of relevant documents in the top-k candidates. R-Precision indicates the proportion of the relevant articles in the top-k ranked ones where k is the exact number of relevant articles for that query. Higher scores in these metrics indicate better performance. 

From Table \ref{main-results}, comparing CuSINeS to the baseline with BM25-based fixed negative sampling strategy across all four models, CuSINeS consistently outperforms the baseline demonstrating its efficacy. This can be attributed to  (i) incorporation of the structural information through hierarchical and sequential proximity to derive the difficulty ranking of negatives. 
(ii) curriculum-based negative schedule rather than providing hard negatives from the start as in baseline. This easy-to-difficult curriculum  helps 
to learn  coarse-grained distinctions between the articles initially and then progressively move towards 
finer-grained nuances. (iii) dynamic criterion of semantic-based difficulty ranking using the model that is being fine-tuned rather than using static BM25 which is based on term matching and independent of the model being pre-trained. 

Overall, dense models outperformed BM25, addressing the lexical gap problem. Legal pre-training and the GNN on top of DR to enrich article representations from the legislative structure 
continued to demonstrate the same trend even with CuSINeS sampling. Notably, CuSINeS demonstrated improvements over GNN-based approaches, highlighting the complementary nature of structure-based sampling compared to structure-based representation learning through graph networks.

\subsection{Ablation Study} 
We study the effectiveness of each of the sub-components in CuSINeS by ablating on DR+GNN (LegalCamemBERT) model.
\newline

\noindent \textbf{Incorporating structural information for negative mining} We use the model with fixed training schedule with BM25-based negatives (baseline). To demonstrate the effect of structural information, we change the negative mining strategy from semantic BM25 distance to structure-based distances computed using sequential and hierarchical views. Further, we also combine these difficulty rankings using RRF. Figure \ref{strcture-fig} demonstrates structure-based negatives are more informative than semantic ones with improvements on R@k at higher k-values. Combining both of them validates their complementary nature with R@k improvement on lower k-values. This result demonstrates that legislation structure information can be leveraged to mine hard negatives to improve SAR performance.
\newline

\noindent \textbf{Training Scheduler} Taking the best model from previous ablation, we change the fixed scheduler to curriculum scheduler for negative sampling with three difficulty buckets where the model witnesses negatives progressively from easy to difficult. Fig. \ref{scheduler-fig} demonstrates that the curriculum-based scheduling achieved better recall indicating this progressive exposure lends the model to find better local optima. This result underscores the need for better ranking criteria to design effective curricula, mirroring the way humans learn, to further enhance model performance in such complex tasks.
\newline

\noindent \textbf{Semantic Difficulty Ranking} We pick the best model from previous setup and ablate the static BM25-based ranking criterion for semantic view with dynamic criterion computed using the being-trained model. From Fig. \ref{semantic-fig}, we observe dynamic semantic ranking criterion helps the model to improve performance, 
illustrating that computing difficulty dynamically helps to design adaptive curricula needed for the model based on its competence at the current training step. 
\newline

\noindent \textbf{Number of Buckets} We ablate on number of buckets which determines the scale of curricula progression. From Fig. \ref{buckets-fig}, we observe that the performance first improves with the increase in number of buckets (upto 3) and then drops as the buckets further increase. This indicates that  a moderate number of buckets strike a balance. 
Too few buckets mix diverse difficulties hindering training, while more buckets narrow focus to specific difficulties, risking forgotten patterns as training advances. 

\section{Conclusion}
We improved the SAR performance 
with CuSINeS, our model-agnostic negative sampling method. It leverages structural information from statutes to assess negative sample difficulty and dynamically update semantic difficulty computed from the model in training. Additionally, curriculum-based training  schedule 
further boosts performance. Our experiments 
on BSARD illustrate each CuSINeS component's contributions, inspiring further research in leveraging legal code structure for enhanced modeling and developing better difficulty assessment methods for curricula design in various legal tasks.


\section{Limitations}
Our experimental contributions are contextual to the Belgian legislation, whose statutes are organized in a topological structure. We believe our negative sampling approach, CuSINeS, is general and could potentially be applied to most jurisdictions with structurally organized statutes. However, the performance of CuSINeS may vary across jurisdictions due to differences in legal nature, semantic difficulty, and statute organization. We leave the exploration of our method in other jurisdictions as future work. Additionally, it is worth noting that constructing such datasets can be expensive and challenging. The BSARD dataset used for evaluation introduces a linguistic bias as Belgium is a multilingual country with French, Dutch, and German speakers, but the provided legal questions and provisions are only available in French \cite{louis2023finding}.

Our work focuses on the first stage of the retrieval system, optimizing for recall. To make this system practically useful, it would require a re-ranking component to sort the retrieved articles by importance, optimizing precision and pinpointing the exact set of statutes needed to answer each question. Furthermore, to achieve the goal of increasing accessibility, the system should not only retrieve relevant articles but also possess the capability to simplify these legal texts, making them more comprehensible to laypeople.

Furthermore, CuSINeS employs curriculum learning to determine the ordering of negative articles for every query during training. Future work can explore understanding query difficulty based on legal complexity, underspecification, or other factors to design even more effective curricula based on queries.

\section{Ethics Statement}
We experiment with a publicly available SAR dataset, BSARD \cite{louis2022statutory}. We are conscious that, by adapting pre-trained encoders, our models inherit any biases they contain and therefore should naturally be scrutinized against applicable equal treatment imperatives regarding their performance, behavior and intended use. Apart from these, we do not foresee any harm incurred by our proposed method.

\section{Bibliographical References}\label{sec:reference}

\bibliographystyle{lrec-coling2024-natbib}
\bibliography{lrec-coling2024-example}


\appendix
\section{Appendix}

\begin{table*}[!ht]
\begin{tabular}{|l|l|l|c|c|c|}
\hline
\textbf{Method}           & \textbf{Encoder Model}                                                     & \textbf{Training Schedule}     & \textbf{R@100} & \textbf{R@200} & \textbf{R@500} \\ \hline
\multirow{3}{*}{DR}       & \multirow{6}{*}{CamemBERT}                                                 & Semantic (Static)              & 77.1           & 81.8           & 86.7           \\ 
                          &                                                                            & Structural                     & 78.8           & 83.1           & 89.3           \\ 
                          &                                                                            & Semantic (Static) + Structural & 77.5           & 84.3           & 89.7           \\ \cline{1-1} \cline{3-6} 
\multirow{3}{*}{DR + GNN} &                                                                            & Semantic (Static)              & 80.2           & 83.2           & 88.6           \\ 
                          &                                                                            & Structural                     & 80.9           & 84.8           & 89.1           \\ 
                          &                                                                            & Semantic (Static) + Structural & 80.8           & 85.2           & 90.4           \\ \hline
\multirow{3}{*}{DR}       & \multirow{6}{*}{\begin{tabular}[c]{@{}l@{}}Legal\\ CamemBERT\end{tabular}} & Semantic (Static)              & 79.8           & 83.9           & 88.9           \\ 
                          &                                                                            & Structural                     & 79.9           & 84.9           & 90.6           \\ 
                          &                                                                            & Semantic (Static) + Structural & 78.7           & 84.4           & 89.6           \\ \cline{1-1} \cline{3-6} 
\multirow{3}{*}{DR + GNN} &                                                                            & Semantic (Static)              & 82.6           & 85.6           & 90.1           \\ 
                          &                                                                            & Structural                     & 82.1           & 86.2           & 91.4           \\ 
                          &                                                                            & Semantic (Static) + Structural & 83.4           & 87.2           & 90.4           \\ \hline
\end{tabular}
\caption{Effect of Incorporating Structural Information on all four models.}
\label{structural-tab}
\end{table*}

\begin{table*}[!ht]
\centering
\begin{tabular}{|l|l|l|c|c|c|}
\hline
\textbf{Method}           & \textbf{Encoder Model}          & \textbf{Training Schedule} & \textbf{R@100} & \textbf{R@200} & \textbf{R@500} \\ \hline
\multirow{2}{*}{DR}       & \multirow{4}{*}{CamemBERT}      & Fixed                      & 77.5           & 84.3           & 89.7           \\ 
&                                 & Curriculum                 & 80.8           & 85.9           & 90.6           \\ \cline{1-1} \cline{3-6} 
\multirow{2}{*}{DR + GNN} &                                 & Fixed                      & 80.8           & 85.2           & 90.4           \\ 
&                                 & Curriculum                 & 82.5           & 87.7           & 91.6           \\ \hline
\multirow{2}{*}{DR}       & \multirow{4}{*}{LegalCamemBERT} & Fixed                      & 78.7           & 84.4           & 89.6           \\ 
&                                 & Curriculum                 & 81.5           & 85.9           & 91.5           \\ \cline{1-1} \cline{3-6} 
\multirow{2}{*}{DR+GNN}   &                                 & Fixed                      & 83.4           & 87.2           & 90.4           \\ 
&                                 & Curriculum                 & 84.2           & 88.8           & 92.6           \\ \hline
\end{tabular}
\caption{Effect of Training Schedule on all four models.}
\label{scheduler-tab}
\end{table*}

\begin{table*}[!ht]
\centering
\begin{tabular}{|l|l|l|c|c|c|}
\hline
\textbf{Method}           & \textbf{Encoder Model}                                                     & \textbf{\begin{tabular}[c]{@{}l@{}}Semantic \\ Negatives\end{tabular}} & \textbf{R@100} & \textbf{R@200} & \textbf{R@500} \\ \hline
\multirow{2}{*}{DR}       & \multirow{4}{*}{CamemBERT}                                                 & Static                                                                 & 80.8           & 85.9           & 90.6           \\ 
&                                                                            & Dynamic                                                                & 82.6           & 86.6           & 91.6           \\ \cline{1-1} \cline{3-6} 
\multirow{2}{*}{DR + GNN} &                                                                            & Static                                                                 & 82.5           & 87.7           & 91.6           \\ 
&                                                                            & Dynamic                                                                & 83.2           & 88.1           & 92.6           \\ \hline
\multirow{2}{*}{DR}       & \multirow{4}{*}{\begin{tabular}[c]{@{}l@{}}Legal\\ CamemBERT\end{tabular}} & Static                                                                 & 81.5           & 85.9           & 91.5           \\
                          &                                                                            & Dynamic                                                                & 83.7           & 87.5           & 92.3           \\ \cline{1-1} \cline{3-6} 
\multirow{2}{*}{DR+GNN}   &                                                                            & Static                                                                 & 84.2           & 88.8           & 92.6           \\ 
&                                                                            & Dynamic                                                                & 84.9           & 89.6           & 93.3           \\ \hline
\end{tabular}
\caption{Effect of Semantic Difficulty Ranking on all four models.}
\label{semantic-tab}
\end{table*}

\subsection{Ablation Study}
\label{ablation}
\textbf{Effect of Incorporating structural information for negative mining:} Tab. \ref{structural-tab} showcases the impact of utilizing structural information on all four models. We create three variants of each model employing difficulty rankings obtained from (a) BM25-based Semantic criterion, (b) Hierarchical- and Sequential-based Structural criterion (c) Combining all of them using RRF. All these variants use fixed training schedule. Table \ref{scheduler-tab} shows that using the structure of statutes yields improvements demonstrating that difficulty assessment can be derived using structural information. Combining both structural and semantic criteria further boosts recall. This highlights the potential of integrating structural insights for more effective negative mining strategies.

\noindent  \textbf{Effect of Training Scheduler:} The results in Table \ref{scheduler-tab} compare two training schedules for each of the four models. One uses a fixed schedule, where the model encounters difficult negatives right from the start of training iterations. The other employs a curriculum schedule, gradually introducing difficult negatives and starting with easier ones. All variants use rankings from three lists based on semantics, hierarchy and sequence. Table \ref{scheduler-tab} shows that the curriculum-based schedule outperforms, indicating that gradually exposing the model to difficulty leads to better results. 

\noindent \textbf{Effect of Semantic Difficulty Ranking :}  In Table \ref{semantic-tab}, we compare two variants of each of the four models based on semantic difficulty ranking. In one variant, the ranking is obtained using BM25 which remains static, and in the other, it's dynamically derived using the model undergoing training. All variants follow a curriculum schedule and use difficulty rankings from three lists. Table \ref{semantic-tab} clearly shows that incorporating dynamic ranking criteria helps improve model performance. This demonstrates that dynamically computing difficulty rankings allows the model to derive its learning curriculum-based on its current training progress, leading to enhanced results.

\end{document}